\documentclass[12pt]{article}
\usepackage{scicite}
\usepackage{times}
\usepackage{graphicx}
\usepackage{pdfpages}

\newenvironment{sciabstract}{%
\begin{quote} \bf}
{\end{quote}}

\newcommand{\bj}{SN\,2002bj}
\newcommand{\bvri}{\protect\hbox{$BV\!RI$} }

\newcommand{\Msol}{M$_{\odot}$}
\newcommand\ion[2]{#1$\;${\scshape{#2}}}
\def\apjl{ApJL }
\def\aj{AJ }
\def\apj{ApJ }
\def\apjl{ApJL }

\def\araa{ARA\&A }

\def\nat{Nature }
\def\iaucirc{IAU Circ. }
\def\mnras{MNRAS}

\newcounter{lastnote}
\newenvironment{scilastnote}{%
\setcounter{lastnote}{\value{enumiv}}%
\addtocounter{lastnote}{+1}%
\begin{list}%
{\arabic{lastnote}.}
{\setlength{\leftmargin}{.22in}}
{\setlength{\labelsep}{.5em}}}
{\end{list}}

\title{An Unusually Fast-Evolving Supernova} 

\author{
Dovi Poznanski,$^{1,2,\ast}$ 
Ryan Chornock,$^{1}$ 
Peter E. Nugent,$^{2}$ \\
Joshua S. Bloom,$^{1}$ 
Alexei V. Filippenko,$^{1}$
Mohan Ganeshalingam,$^{1}$ \\
Douglas C. Leonard,$^{3}$ 
Weidong Li,$^{1}$ and
Rollin C. Thomas$^{2}$\\
\small{$^{1}$Department of Astronomy, University of California, Berkeley, CA 94720-3411}\\
\small{$^{2}$Computational Cosmology Center, Lawrence Berkeley National Laboratory,}\\
\small{ 1 Cyclotron Road, Berkeley, CA 94720}\\
\small{$^{3}$Department of Astronomy, San Diego State University, Mail Code 1221, San Diego, CA 92182-1221}\\
\small{$^\ast$To whom correspondence should be addressed; e-mail:  dovi@berkeley.edu}
}

\date{}

\begin{document} 
\baselineskip24pt
\maketitle 
\begin{sciabstract}
Analyses of supernovae (SNe) have revealed 
two main types of progenitors: exploding white dwarfs and collapsing massive stars. 
We present \bj, which stands out as different from 
any SN reported to date. 
Its light curve rises and declines very rapidly, yet reaches a peak intrinsic
brightness greater than $-{\bf 18}$\,mag. A spectrum obtained 7 days
after discovery shows the presence of helium and intermediate-mass
elements, yet no clear hydrogen or iron-peak elements. 
The spectrum only barely resembles that of a Type Ia supernova, with added carbon and helium.
Its properties suggest that \bj\ may be representative of a class of progenitors that previously has been only hypothesized:
a helium detonation on a white dwarf, ejecting a small envelope of material.
New surveys should find many such objects, despite their scarcity.

\end{sciabstract}

 Supernovae (SNe) are usually classified based on tell-tale lines in
 their spectra \cite{filippenko97}.  Those empirical types are routinely associated with
 progenitor systems and the understanding of their explosion
 mechanisms.  Type Ia SNe are interpreted as the thermonuclear
 disruption of a white dwarf, and the other types as the core collapse
 of a massive star.  \bj, which we present here, would formally belong according
 to that classification to the Type Ib class, due to the lack of
 hydrogen and the presence of helium in the optical spectra we have
 obtained. However, the overall observed properties of this SN are 
 unprecedented, and the taxonomic classification is misleading.

\bj\ was discovered independently at mag 14.7 by the Lick Observatory
SN Search (LOSS) and by amateur astronomers \cite{puckett02}, on
2002 Feb. 28.2 (UT dates are used throughout this paper), in the galaxy 
NGC\,1821. The distance corrected for local bulk flows (assuming
$H_0=73$\,km\,s$^{-1}$Mpc$^{-1}$) is $50\pm5$\,Mpc (see SOM for a
discussion of the host-galaxy properties and distance).  A pre-discovery LOSS
image with limiting magnitude 18.4 (Galactic extinction corrected), 
on 2002 Feb. 21.2 (a week before discovery), shows
nothing at that position. 

As part of our SN follow-up program, we obtained
\bvri\ photometry of \bj\ for 9 epochs over 20 days until it faded
below the detection threshold (SOM). Our photometry does not show a rising
phase, but the non-detection constrains the rise to be of  less than 7
days. The decline is almost as fast, dropping by 4.5 mag (in the $B$
band) in 18 days. \bj\ evolves on unprecedented timescales (Fig. 1).

The spectrum we obtained a week after detection is extremely blue, with
weak, yet remarkable, features \cite{note1}. Using a $\chi^2$ fit, we have digitally compared our (continuum-removed) spectrum with about 4000 spectra of nearly 1400 SNe, allowing for velocity offsets. Not a single spectrum fits well.  The closest matches were SNe\,Ia, mostly due to the absorption feature near 6150\,\AA\ (rest frame), usually attributed to \ion{Si}{II} (Fig. 2). 
The best of
those (SN\,2009dc) is a superluminous, slowly declining, carbon-rich,
possibly super-Chandrasekhar-mass SN\,Ia
\cite{howell06,hicken07,yamanaka09}.  These few very luminous SNe reported
so far evolve slowly and eject significant
amounts of unburned material, suggesting massive white dwarf progenitors.
That is, the closest spectroscopic match has one of the most substantially different light curves. While
the spectra are broadly similar, \bj\ has prominent \ion{He}{i}
lines, which are not expected in a SN\,Ia.  In addition, the spectrum
of SN\,2009dc had to be artificially redshifted by 3,000\,km\,s$^{-1}$
in order to match that of \bj, implying that \bj\ had slower ejecta at the time the spectrum was taken.

Using the code SYNOW \cite{fisher97} we produced synthetic spectra and
identified most of the features as coming from helium and
intermediate-mass elements such as carbon, silicon, and sulfur, but no hydrogen (see SOM). While this empirical fit does not produce
meaningful abundances, the lack of iron or other iron-peak
elements in the fit is peculiar. As exceptional is the considerable
\ion{S}{ii} contribution, when compared to \ion{Ca}{ii}  
(a ratio never seen before in other SNe). We also
report a tentative identification of \ion{V}{ii}. While based on
only a single line, the relevant spectral region ($\sim$3950\,\AA)
would have emission from \ion{Ca}{ii} without it.
The spectrum was taken in spectropolarimetry mode, yet there are
 no polarization line features down to 0.1--0.2\%. The continuum is
consistent with polarization by dust in the Milky Way (see SOM for details).

Using our photometry we determined the bolometric evolution of
\bj\ (SOM). The total radiated energy in optical bandpasses
is of order $10^{49}$\,erg, 
starting at a peak of $10^{43}$\,erg\,sec$^{-1}$. 
Assuming blackbody emission we derived the temporal evolution
of the effective temperature, radius, and photospheric velocity.  
The temperature and velocity decline very rapidly, indicating 
rapid recession of the photosphere in a low-mass envelope.  We
estimated the mass of the ejecta using the scaling
relation that ties it to the photospheric velocity and rise time,
$M_{\rm ej,1}=\left(\frac{t_1}{t_2}\right)^2 \frac{v_1}{v_2}M_{\rm ej,2}$
\cite{arnett82}.  This scaling assumes the opacity is similar to that of a SN\,Ia. \bj\ rises at least 3 times faster than a normal
SN\,Ia (depending on the assumed explosion date); 
thus, while its velocity at peak is uncertain (see discussion
in SOM), the ejected mass has to be smaller than $\sim$0.15\,M$_{\odot}$, about 10\% that of a SN\,Ia. 

The luminosity and short rise time of \bj\ translate to 0.15--0.25\,\Msol\ of
$^{56}$Ni when using Arnett's law \cite{arnett82,sutherland84}, if the
light curve is solely powered by radioactive $^{56}$Ni and its decay product $^{56}$Co.  Under this
same assumption, the rapid decline we measured requires a sharp drop of the
gamma-ray deposition efficiency of an order of magnitude in less than
3 weeks.

The small ejected mass we derived, the lack of iron-peak elements in the spectrum, and the 
relatively large amount of $^{56}$Ni required to explain the high luminosity are difficult to reconcile
without summoning an additional energy source.

\bj\ looks spectroscopically
somewhat like a SN\,Ia but with helium, carbon, and an exceptionally fast light curve.  
Recently, a mechanism has been proposed \cite{bildsten07} by which
binary white dwarfs of the AM CVn class may undergo a thermonuclear
explosion of the helium accreted on the primary star. Such a scenario
will produce roughly 10\% of the luminosity of a SN\,Ia, for about 10\%
of the typical time --- hence these objects were dubbed ``.Ia'' supernovae. 

\renewcommand{\labelitemi}{$-$}
	\begin{itemize}
\item These SNe are expected to be faint (between $-15$ and
  $-18$ mag at peak in the $V$ band) and rapidly evolving (1--6 days
  of rise time, with the brighter objects usually rising more
  slowly). While the decline was not explicitly discussed by Bildsten et al. \cite{bildsten07}, the low ejected mass implies a rapid decline.

\item The short timescales of these events may allow the detection of the short-lived radioactive
  nuclei, $^{52}$Fe or $^{48}$Cr, in addition to the standard
  $^{56}$Ni which drives SN\,Ia light curves. $^{48}$Cr decays to $^{48}$V within a day, and then to 
$^{48}$Ti in a week. The decay of these nuclei may (partially) power the optical light curve.

\item The rate of .Ia events is predicted to be roughly a few percent
  of the SN\,Ia rate per unit local volume.
\end{itemize}

While the spectral signature was not predicted, some properties 
seem to result naturally in that scenario. Because this is a thermonuclear
helium detonation on a white dwarf, we do not expect any hydrogen, but helium
does seem reasonable, as well as intermediate-mass elements that
either survive the convective burning phase and detonation \cite{shen09} or are produced in the explosion. 

While other recent supernovae have been proposed to be related to helium detonations on a WD [SN\,2005E, \cite{perets09} and 
perhaps also SN\,2008ha \cite{perets09,valenti09,foley09}], these
events have more massive ejecta ($0.2-0.3\,M_{\odot}$) and much slower light 
curves, and thus do not fit the current predictions of .Ia models, though
they may be explained with related phenomena involving much more massive
helium shells.

The light curve of \bj\ is as fast as predicted in this model, but
slightly more luminous than expected \cite{note2}. 
The high luminosity yet small ejecta mass may be reconciled if some
short-lived $^{48}$Cr or $^{52}$Fe are synthesized, as their yield per unit mass is higher
than that of $^{56}$Ni on short timescales.
Our tentative identification of \ion{V}{ii} in the spectrum supports
this hypothesis ($^{48}$V is the daughter of $^{48}$Cr), 
and the peculiar composition of the spectrum may as
well. SNe\,Ia usually display a prominent secondary peak in their infrared light curve, 
attributed to line blanketing by singly ionized iron-peak elements \cite{kasen06}. 
The lack of a secondary peak in the light curve of \bj\ is consistent, within that picture, 
with our non-detection of iron-peak elements in the spectrum.  

Bildsten et al. \cite{bildsten07} assumed that the ejecta will have
velocities of $\sim$15,000\,km\,s$^{-1}$, which is equivalent
to all the binding energy released from fusing helium converted into
kinetic energy (as the star is assumed to be left bound). 
We infer for \bj\ photospheric velocities that drop rapidly from
about 8400\,km\,s$^{-1}$ at detection to 2,000\,km\,s$^{-1}$ three weeks later. Extrapolating to an explosion date 7 days before detection, 
the initial
velocity could have been between 14,000\,km\,s$^{-1}$ (linear
extrapolation) and $\sim$25,000\,km\,s$^{-1}$ [exponential
extrapolation as often seen in SNe \cite{nugent06,wang09a}]. 
A rise time which is faster by a factor of 2 would 
imply velocities twice as high. The only direct measurement we have is from the
spectrum, 7 days past detection, about 4000\,km\,s$^{-1}$. This is consistent with the
derived photospheric velocity at that time if the rise time was about 7 days.

Out to a distance of 60\,Mpc, the LOSS survey is complete (99\%) for SNe\,Ia and 31
have been found. Because \bj\ is quite luminous, the incompleteness
correction for it is almost as small (94\%), resulting in a
relative rate of 3.4\% of the SN\,Ia rate for \bj-like SNe \cite{note3}. 
This is in good agreement with the predictions for SNe\,.Ia.

The SN\,.Ia model, still in its infancy, lacks more stringent
predictions such as detailed light curves and the spectral
composition and evolution. Nevertheless, all the diagnostics we could
apply seem consistent with this interpretation. 
While the evidence here is tentative, the existence of vanadium, 
if seen in future discoveries of objects of this class, points to
a different nucleosynthetic chain and therefore may serve as a
smoking gun for a truly different supernova explosion channel. 
Regardless of the interpretation,  current and future surveys should focus 
on short cadences -- repeat visits on daily rather 
than weekly timescales ---  in order to find many more SNe resembling \bj.

\begin{scilastnote}
	
\item We thank L. Bildsten for valuable insights into the SN~.Ia model, A. Gal-Yam, D. Kasen, D. Maoz, T. Matheson, P. Mazzali, E. Ofek, E. Quataert, K. Shen, and N. Smith  for useful discussions, R. Foley for reducing the Lick 3-m spectrum of SN\,2002bj, and A. A. Miller and A. Merritt for the DeepSky analysis.
A.V.F.'s group has been supported by US National Science
  Foundation (NSF) grants AST--0607485 and AST--0908886, by Department
  of Energy grants DE-FC02-06ER41453 (SciDAC) and DE-FG02-08ER41563,
  and by the TABASGO Foundation.  KAIT and its ongoing operation were
  made possible by donations from Sun Microsystems, Inc., the
  Hewlett-Packard Company, AutoScope Corporation, Lick Observatory,
  the NSF, the University of California, the Sylvia \& Jim Katzman
  Foundation, and the TABASGO Foundation.  Some of the data presented
  herein were obtained at the W. M. Keck Observatory, which is
  operated as a scientific partnership among the California Institute
  of Technology, the University of California, and the National
  Aeronautics and Space Administration; the Observatory was made
  possible by the generous financial support of the W. M. Keck
  Foundation. We thank the staffs at the Lick and Keck Observatories
  for their assistance.
\end{scilastnote}

\clearpage
\noindent \includegraphics[width=6.25in]{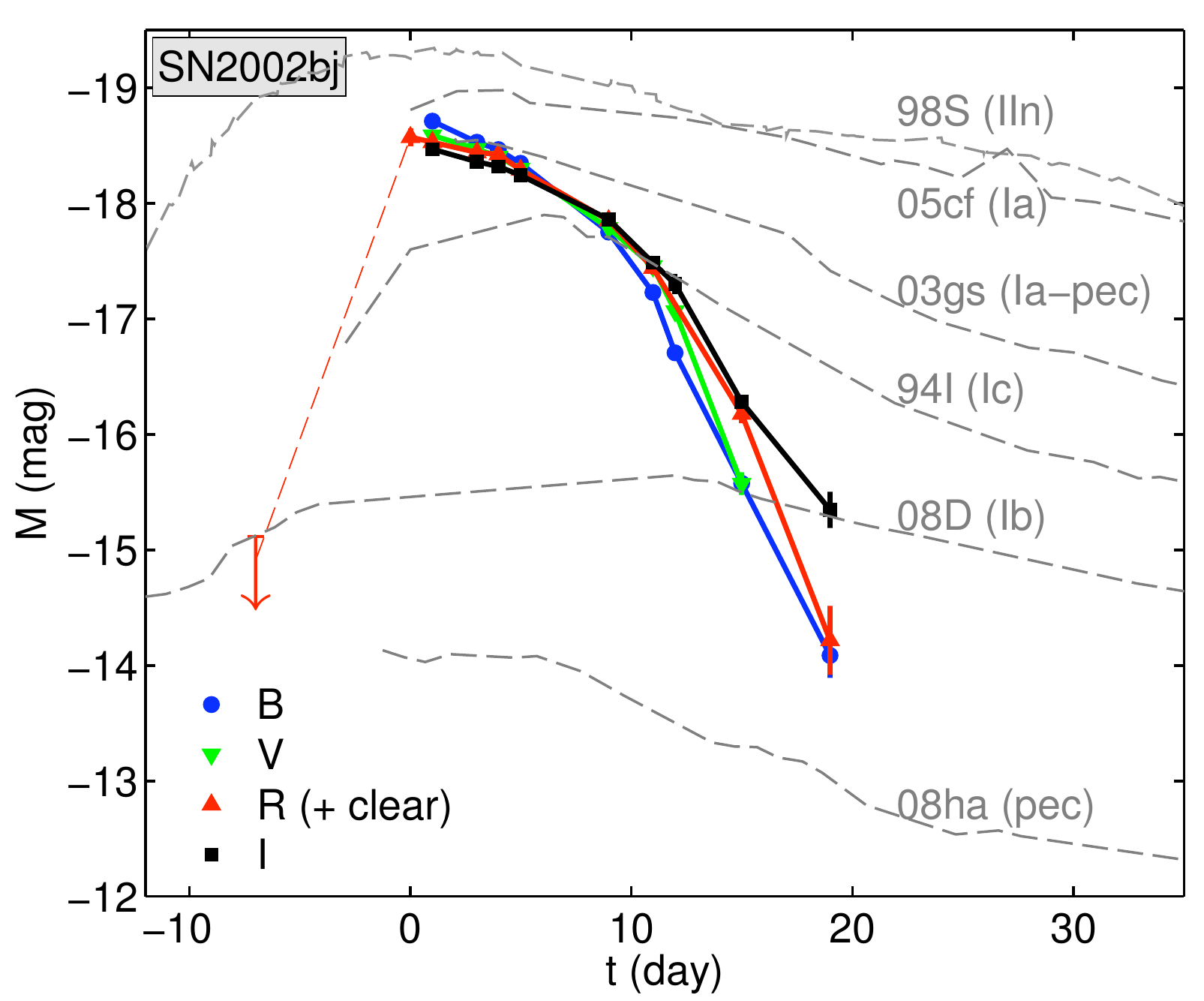}\\
\noindent {\bf Fig. 1.} Comparison of the light curve of \bj\ to those of SNe of
various types (in the $R$ band offset to the same $B$-band maximum date).  \bj\
 is quite luminous
at peak for a core-collapse event, yet faint compared with typical
SNe\,Ia. 
SN\,1994I \cite{richmond96} is often cited as a ``fast'' SN Ic,  
SN\,2003gs \cite{krisciunas09a} was recently presented as one of the fastest SNe\,Ia,
and SN\,2008ha \cite{foley09} is a faint, peculiar, and fast SN of debated breed. \bj\ is significantly faster
than any of these. 
SN\,1998S \cite{fassia00}, SN\,2005cf \cite{wang09}, and SN\,2008D \cite{modjaz09} 
are standard representatives of type IIn, Ia, and Ib SNe, respectfully; they are shown for reference. The dashed red line shows the {\it slowest} rise slope of SN\,2002bj allowed by the data. 

\clearpage
\noindent \includegraphics[width=6.25in]{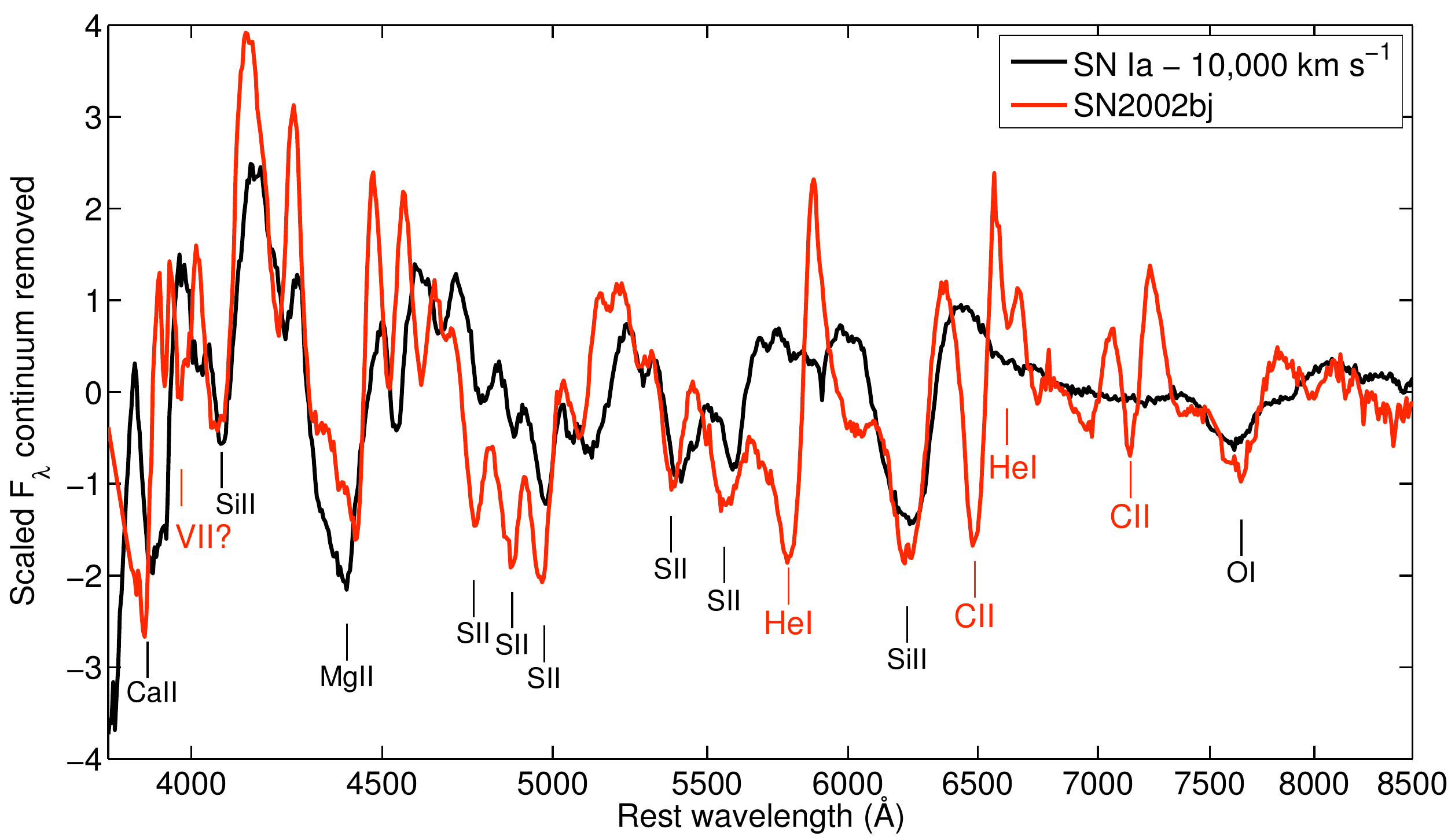}\\
\noindent {\bf Fig. 2.} 
The unique spectral features of \bj\ (shown in red; continuum removed) are difficult to identify  a priori. This spectrum, taken March 7, 2002  (7 days after discovery), is reminiscent of SNe\,Ia, with the notable exception of the prominent helium and carbon lines,
never seen before in such SNe. We show (in black) a typical SN\,Ia spectrum near maximum light (SN\,2001bf), redshifted by 10,000\,km\,s$^{-1}$ in order to match ejecta velocities.  The spectral features identified in black are present in both objects and the ones in red are seen only in \bj.
\\
\includepdf[pages=-]{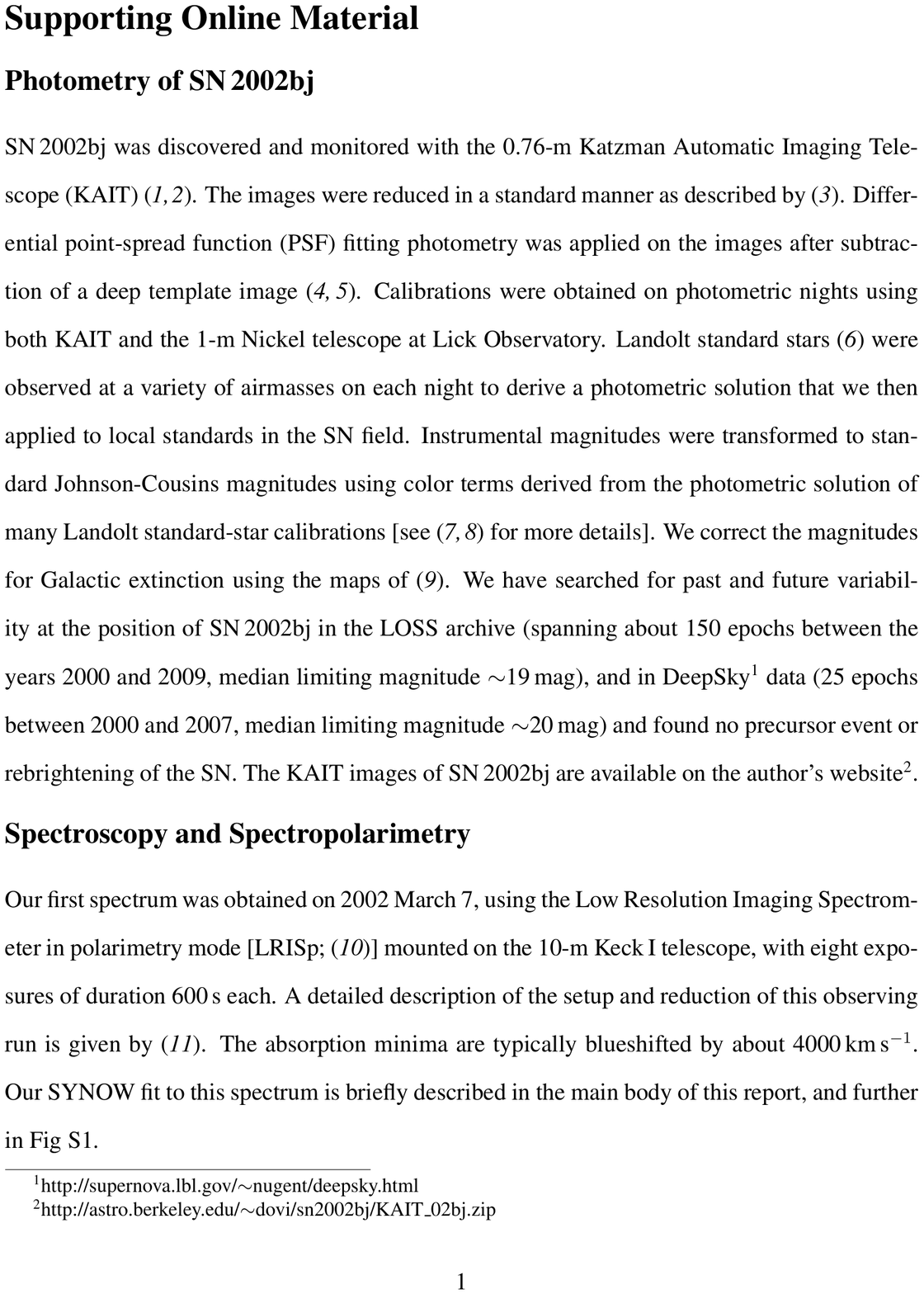}

\begin{thebibliography}{10}
\bibitem{note2} {If our interpretation is correct, it would not be too surprising
that the first SN~.Ia found would be unusually luminous, since such
an object would be easier to find and to recognize as such.}
\bibitem{note3} {Note that Poisson statistics allow fractions in the range
  0.7--11\% (1$\sigma$).}
\end{thebibliography}
\end{document}